\documentclass[epsf,rotating,doublespacing]{elsart}
\usepackage{epsfig}
\usepackage{subfigure}
\usepackage{graphicx}
\usepackage{epstopdf}
\begin{document}
\newcommand{\be}{\begin{equation}}
\newcommand{\ee}{\end{equation}}
\newcommand{\ve}{\varepsilon}

\begin{frontmatter}

\title{Fractals and dynamical chaos \\ in a random 2D Lorentz gas with sinks}
\author{I. Claus,$^1$ P. Gaspard,$^1$ and H. van Beijeren$^2$}
\address{$^1$ Center for Nonlinear Phenomena
and Complex Systems, Facult\'e des Sciences,\\
Universit\'e Libre de Bruxelles, Campus Plaine, Code Postal 231,\\
B-1050 Brussels, Belgium\\
$^2$ Institute for Theoretical Physics, Utrecht University,\\
Leuvenlaan 4, 3584 CE Utrecht, The Netherlands}
\begin{abstract}
Two-dimensional random Lorentz gases with absorbing traps are considered
in which a moving point particle undergoes elastic collisions on hard
disks and annihilates when reaching a trap. In systems of finite spatial
extension, the asymptotic decay of the survival probability is exponential
and characterized by an escape rate $\gamma$, which can be related to the
average positive Lyapunov exponent and to the dimension of the fractal
repeller of the system. For infinite systems, the survival probability
obeys a stretched exponential law of the form $P(c,t)\sim\exp(-Ct^{1/2})$.
The transition between the two regimes is studied and we show that,
for a given trap density, the non-integer dimension of the fractal repeller
increases with the system size to finally reach the integer dimension of
the phase space. Nevertheless, the repeller remains fractal.
We determine the special scaling properties of this fractal.
\end{abstract}

\end{frontmatter}
\section{Introduction}
Many recent works have been devoted to the problem of diffusion in the
presence of static absorbing traps, or sinks. In such models,
the diffusing particles can be seen as modeling some active chemical species
or some kind of excitation. The trapping
then consists in a reaction with another chemical species leading to a
non-reactive product, or in an excitation trapping converting the energy
to some other form \cite{HavBen}. The important quantity to study is the
survival probability $P(c,t)$, i.e., the probability for a
diffusing particle to survive in the system at time $t$ for
a trap concentration $c$. In the case of randomly distributed traps,
in a system of infinite spatial extension, the survival probability
$P(c,t)$ of the diffusing particles obeys asymptotically a stretched
exponential
law of the form \cite{DV,GP}
\begin{equation}
P(c,t) \sim \exp\left[ - A_d \; u^{2/(d+2)} \; t^{d/(d+2)} \right] ,
\label{stretch.exp}
\end{equation}
where $d$ is the dimension of the system, $c$ is the concentration of
static traps, $u=-\ln(1-c)$ and $A_d$ is a constant depending on the dimension
and on the geometry of the system. This behavior is due to the existence
with a non-vanishing probability of very large trap-free regions in which
particles can survive for a very long time \cite{GP}.

The stretched exponential law (\ref{stretch.exp}) has been much studied in
stochastic systems although, until now, no study of this phenomenon
has been carried out in deterministic chaotic systems.  Such a study is
of special interest in deterministic chaotic systems where the trapping process
is characterized not only by the behavior of the survival
probability but also by the fractal set of all the trajectories
that never meet a trap.  This set is defined in the phase space
of the deterministic system and is referred to as a fractal repeller
\cite{GB95,G98}.

The fractal repeller associated with the deterministic chaotic diffusion
in the presence of {\it periodically} distributed traps or sinks
has been studied in a previous paper \cite{CG01}.
The system studied there was a two-dimensional periodic Lorentz gas:
a point particle undergoes elastic collisions on hard disks fixed in the
plane and forming a regular triangular lattice. Some of these disks, forming a
regular superlattice over the disk lattice, play the role of traps: when
the point particle collides on one of them, it is annihilated.
In this case, since the size of trap-free regions is bounded, the decay
is exponential, characterized by an escape rate $\gamma$, so
\begin{equation}
P(c,t) \sim e^{-\gamma t}.
\label{esc}
\end{equation}
The dynamics is chaotic and the repeller has a non-integer fractal dimension.

The purpose of the present paper is to introduce a model for the
stretched exponential law (\ref{stretch.exp}) which is based on
a deterministic dynamics in order to be able to investigate the
deterministic features of the process and, especially, the properties
of the associated fractal repeller.  We might suspect that the
fractal dimension of the repeller is integer and
equal to the phase space dimension $D_H=3$ since, in a random system
of infinite spatial extension, the survival probability (\ref{stretch.exp})
decays more slowly than the exponential (\ref{esc}).  However,
if the dimension is integer, a finer characterization of the repeller
is required and we might even wonder if the repeller is still a fractal.

In order to find a solution to this puzzle, we first apply
the escape-rate formalism to {\it finite} two-dimensional random Lorentz
gases.
We confirm that, in such systems, the asymptotic decay is exponential
and that the corresponding escape rate can be related to the positive Lyapunov
exponent and to the fractal dimension of the fractal repeller.

We next consider {\it infinite} 2D random Lorentz gases.
Thanks to our fast event-driven numerical algorithm
and appropriate simulation method, we are able
to extend the study of the survival probability to these
deterministic chaotic systems.

Finally, we study the transition from finite to infinite
2D random Lorentz gases, which then allows us to determine
the evolution of the scaling properties of the fractal
repeller when the system size increases and to solve the
aformentioned puzzle.

The paper is organized as follows. Section \ref{ranfin} is devoted to
finite two-dimensional random Lorentz gases with sinks. The
asymptotic decay is shown to be exponential, described by an escape rate
$\gamma$. Moreover, we show that this open chaotic system is characterized
by the existence of a fractal repeller. In these open systems,
the average Lyapunov exponent and the fractal dimensions of the repeller
are calculated with a non-equilibrium probability measure. The validity of the
relation (\ref{esc}) is confirmed for these finite size systems.
The case of infinite two-dimensional random Lorentz gases is
presented in section \ref{raninf}.
Thanks to our appropriate simulation method,
the stretched exponential decay is observed for the first time in a
deterministic chaotic system. The transition from finite systems
with exponential decay to infinite ones obeying stretched exponential laws
is studied in section \ref{rantrans}. For a given concentration of traps,
the Hausdorff dimension of the repeller is shown to increase with
system size, to finally reach an integer value equal to the dimension of
phase space, as it would be the case for a closed system.
We show that, nevertheless, the repeller remains fractal
in the infinite-size limit and that it is no longer characterized
by a simple power-law scaling as the classical fractals but by
a more complicated scaling.
Conclusions are drawn in section \ref{concl}.

\section
{The two-dimensional random Lorentz gas of finite spatial extension}
\label{ranfin}
\subsection{Description of the model}

The 2D random Lorentz gas consists
of a point particle colliding
elastically on randomly
and uniformly distributed non-overlapping
hard disks fixed in the plane.
In finite (or periodic) geometry we will require in addition
that the horizon is finite, i.e., that
the point particle, starting from any initial condition,
collides on a disk after a finite time.
Under these conditions the system is
diffusive, with a well-defined diffusion coefficient. Some disks are
chosen to be sinks: when the point particle collides on one of them, it is
absorbed, according to the following reaction scheme:
\begin{eqnarray}
{\rm X}\ + \ {\rm inert \ disk} \ &\leftrightarrow& \ {\rm X}\
+ \ {\rm inert \ disk} \ ,\\
{\rm X}\ + \ {\rm sink} \ &\rightarrow& \ \emptyset \ + \ {\rm sink} \ .
\end{eqnarray}

In practice, we consider a square unit cell of size $L \times L$.
In this cell, $n_d$ disks of unit radius are placed randomly
in the following way.
The disks are first placed periodically, on a square or triangular
lattice.
A random velocity of unit magnitude is assigned to each of them,
and the system is evolved as a hard ball system
during a relaxation time $T_{relax}$. The
configuration is then frozen and $n_s$
disks are randomly chosen to be sinks.
Each disk has the probability $c$ to be a sink,
independently of the others, so $n_s$ is always close to $cn_d$.
Three examples of such unit cells are shown in Fig.\ \ref{unit12}.

The motion of the point particle in the system is defined
under periodic
boundary conditions. The collisions on the disks are elastic so that the
energy of the system is conserved.
The magnitude of the point particle velocity is therefore a
constant, fixed here to one. The particle velocity can
be specified by a unique coordinate, for example the angle $\psi$ between
the
velocity and the $x$-axis. Its position is specified by two coordinates
$\{x,y\}$. The three phase-space variables of the point particle are
therefore $\Gamma=\{x,y,\psi\}$.

\subsection{Escape and fractal repeller}
\subsubsection{Asymptotic behavior}

To study the asymptotic time behavior in such a system, we define the survival
probability $P(c,t)$, which is the probability for a point particle to
be still in the system at time $t$, for a sink concentration $c$
now taken to be
exactly $n_s/n_d$. For finite systems in fact this survival probability depends
on the positions of the scatterers as well as on the distribution of the
traps, but as system size increases the fluctuations of $P(c,t)$ around its
average
value will become increasingly smaller.
Suppose the long time behavior of $P(c,t)$
can be written in the form
\begin{equation}
P(c,t) \sim \exp\left[-C\;t^{\alpha(t)}\right].
\end{equation}
In the case of an exponential decay, $\alpha$ would be equal to $1$, and
for the stretched exponential behavior
we expect to hold asymptotically in two dimensions, it would be equal to
$0.5$. This may be generalized into a time dependent $\alpha(t)$
defined as \cite{An84}
\begin{equation}
\alpha(t)=\frac{d}{d \ln t}\ln[-\ln P(c,t)].
\label{alpha2}
\end{equation}
This quantity, obtained numerically, is plotted versus time in Fig.\
\ref{alpha12}.

\begin{figure}[htb]
\centerline{\epsfxsize=14 truecm \epsfbox{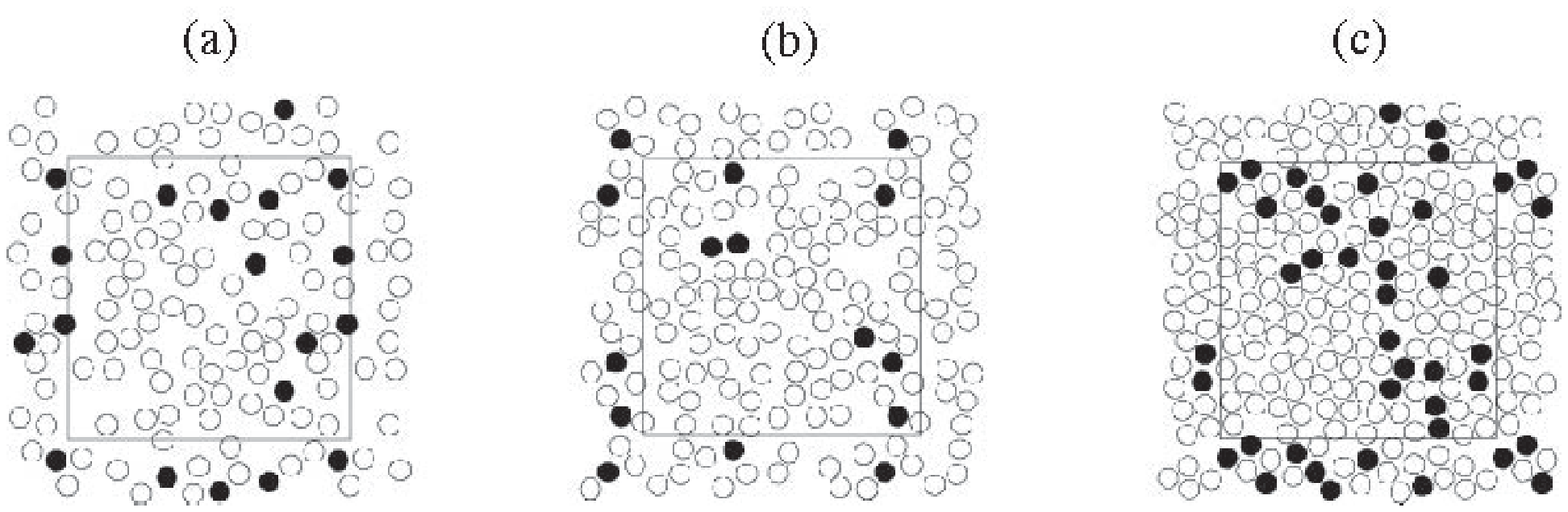}}
\caption{Typical unit cell of a finite two-dimensional random Lorentz gas:
(a) $L=25$, $n_d=72$, $n_s=9$;
(b) $L=25$, $n_d=88$, $n_s=7$;
(c) $L=25$, $n_d=127$, $n_s=23$.
The filled disks are
the sinks.
\label{unit12}}
\end{figure}

\begin{figure}[htb]
\begin{center}
\epsfxsize=6cm
\subfigure[]{\epsfbox{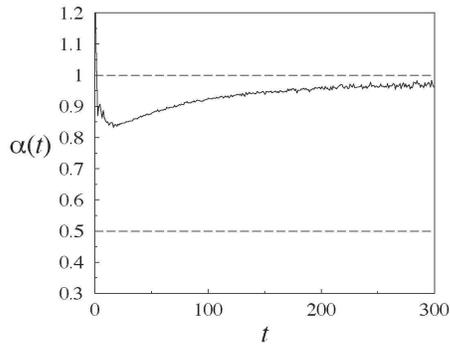}}
\hfill
\epsfxsize=6cm
\subfigure[]{\epsfbox{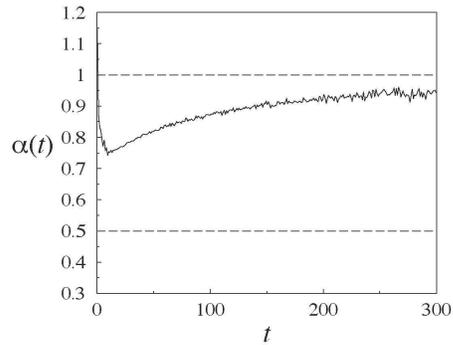}}\\
\epsfxsize=6cm
\subfigure[]{\epsfbox{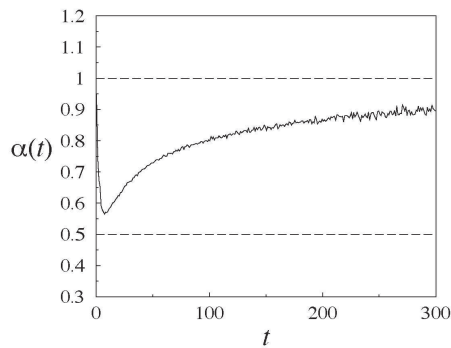}}
\caption{Two-dimensional random Lorentz gas of finite spatial extension;
Effective exponent $\alpha$ plotted versus time, for a trap density
$c \approx 0.05$:
(a) in the case of the disk configuration represented in Fig.\
\ref{unit12}a,
but with $n_s=4$ traps;
(b) in the case of the disk configuration represented in Fig.\
\ref{unit12}b,
but with $n_s=5$ traps;
(c) in the case of the disk configuration represented in Fig.\
\ref{unit12}c,
but with $n_s=6$ traps. \label{alpha12}}
\end{center}
\end{figure}

The first graph, in Fig.\ \ref{alpha12}a, represents the values of
$\alpha$
for the disk configuration represented in Fig.\ \ref{unit12}a,
but with only 4 traps,
corresponding to a trap density $c \approx 0.05$. An average has been
taken over 10 different trap configurations. Figure \ref{alpha12}b
(respectively \ref{alpha12}c) presents the results obtained for the disk
configuration of Fig. \ref{unit12}b (respectively \ref{unit12}c), with 5
(respectively 6) traps, corresponding to $c \approx 0.05$. In
all three cases, we observe that, after a transient, $\alpha$ slowly
tends to 1 \cite{Bunde97}.

\begin{figure}[htb]
\begin{center}
\epsfxsize=6cm
\subfigure[]{\epsfbox{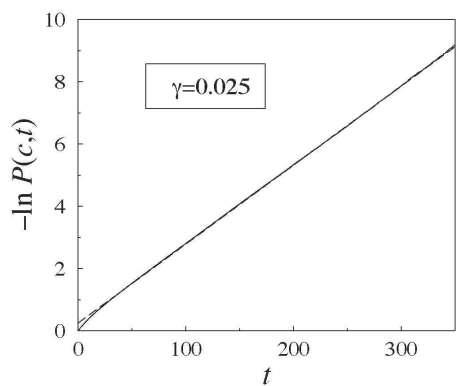}}
\hfill
\epsfxsize=6cm
\subfigure[]{\epsfbox{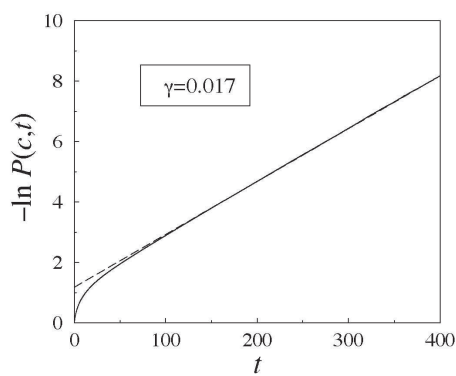}}
\caption{ Two-dimensional random Lorentz gas of finite spatial extension;
Logarithm of the survival probability as a function of time: (a)
for the disk and trap configuration represented in Fig. \ref{unit12}a,
(b) for the disk and trap configuration represented in Fig. \ref{unit12}c.
After a transient time the behavior is exponential and the slope of the
curve gives the value of the escape rate $\gamma$.
\label{esc12}}
\end{center}
\end{figure}

The escape rate characterizing the exponential decay
may be obtained as follows \cite{GB95,G98,CG01}.
Let us take $N_0$ initial conditions randomly distributed
in the phase space of variables $\Gamma=(x,y,\psi)$ according to
some initial measure $\nu_0$.
After a time $t$, only a number $N_t$ of particles will remain in the system.
This number will decrease monotonically
with $t$. The set of particles remaining in the random
Lorentz gas at time $t$
consists of those colliding on a trap, and therefore
escaping, at an {\em escape time} ${\mathcal T}(\Gamma_0)$, which is
larger than $t$:
\begin{equation}
\Upsilon
(t)=\{\Gamma_0 : t < {\mathcal T}
(\Gamma_0)\}.
\end{equation}
The decay of the number of particles is expected
to be asymptotically exponential due to the finite
size of the system. The escape rate is thus defined as
\begin{equation}
\gamma = \lim_{t \to +\infty} -\frac{1}{t}\ln \frac{N_t}{N_0}
=\lim_{t \to +\infty} -\frac{1}{t}\ln \nu_0[\Upsilon (t)] .
\label{esc2}
\end{equation}
The escape rate is a characteristic quantity of the system, independent of
the choice of the initial measure $\nu _0$ as long as this latter
is smooth enough.
The escape rate is easily accessible by numerical computations, as shown
in Fig.\ \ref{esc12}. The logarithm of the survival probability is plotted
versus time in Fig.\ \ref{esc12} for the disk and trap configurations
presented in Figs.\ \ref{unit12}a and c.
The behavior is clearly exponential after a short transient. The slopes
of the curves give escape rates of $\gamma=0.025$ and $\gamma=0.017$
respectively.

\subsubsection{Fractal repeller and nonequilibrium probability measure}
A two-dimensional random Lorentz gas of finite spatial extension
with traps is an open system, obeying an asymptotically
exponential decay. Moreover, its dynamics
is chaotic. It can therefore be studied by the methods developed in
Refs. \cite{GB95,G98,CG01}, in order to relate its microscopic chaotic
dynamics to its macroscopic exponential decay.

An important property of such a system is the existence of a fractal
repeller. For a typical initial condition $\Gamma_0$,
the particle will collide on a trap and escape after a finite escape time
${\mathcal{T}} (\Gamma _0)$.
The same will happen if we follow the trajectory in the negative time
direction. Although
a forward and a backward escape time exist for almost all
trajectories, there
do exist trajectories that never collide on a sink and
remain trapped forever in the Lorentz gas in both time directions.
They can be both periodic and non-periodic.
Because of the defocusing character of the collisions on the disks, these
non-colliding
trajectories are unstable and form a fractal set of zero Lebesgue measure
in phase space: this set is
called the fractal repeller ${\mathcal F}$.
It is a typical feature of chaotic-scattering processes.

\begin{figure}[htb]
\begin{center}
\epsfxsize=6cm
\subfigure[]{\epsfbox{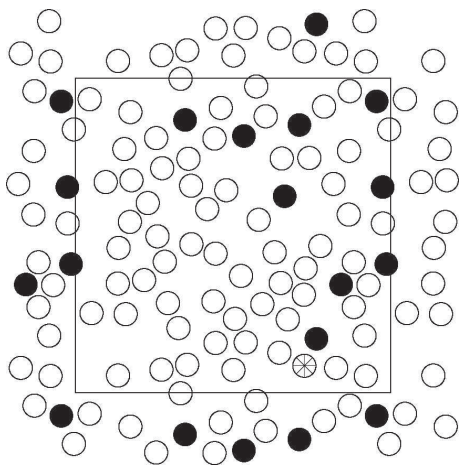}}
\hfill
\epsfxsize=6cm
\subfigure[]{\epsfbox{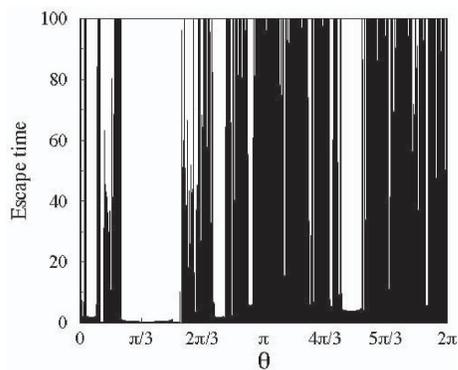}}\\
\epsfxsize=6cm
\subfigure[]{\epsfbox{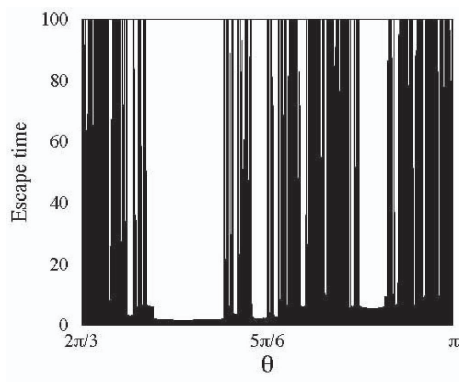}}
\hfill
\epsfxsize=6cm
\subfigure[]{\epsfbox{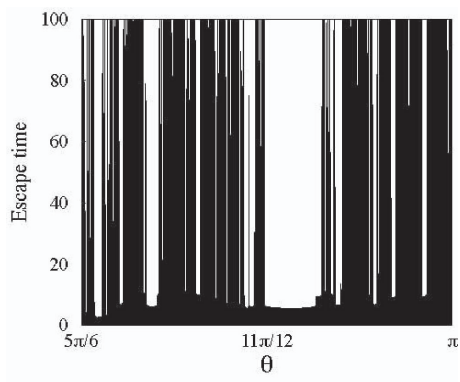}}
\caption{Escape time as a function of the initial position: (a) The
initial
positions are taken at a varying angle $\theta$ around the disk marked by
a
star; the initial velocity is always normal to the disk. Escape time for
(b) $\theta \in [0,2\pi]$; (c) $\theta \in [2\pi/3,\pi]$;
(d) $\theta \in [5 \pi/6,\pi]$.
\label{repeller}}
\end{center}
\end{figure}

Evidence of the fractal character of this repeller is given by the forward
escape
time ${\mathcal{T}} (\Gamma _0)$ as a function of the initial condition,
as shown in Fig.\ \ref{repeller}.
This function is finite for almost
all initial conditions since the particle typically collides
with a sink after a finite time. However,
it is infinite for trajectories trapped on the fractal repeller.
These trajectories start from initial points on the stable manifold
(in the forward direction) of a
trajectory that is trapped in both time directions. The
infinities of the escape-time function thus establish
a fractal set formed by the stable manifold of the fractal
repeller, $W_{\rm s}({\mathcal F}_n)$.

Fig.\ \ref{repeller}b-d shows this escape time as a function of the
initial position, given by an angle $\theta_0$ around the disk marked by a star
in Fig.\ \ref{repeller}a. The velocity is chosen normal to this disk. Due
to the
random distribution of disks, the fractal set is not quasi self-similar,
as it was the case in Ref.\ \cite{CG01}. However,
Figs.\ \ref{repeller}c-d show that a fine structure of escape windows
exists at smaller and smaller scale.
The characteristics of this structure, such as correlation functions and
the distribution of sizes of empty windows, may be expected to
show self-similarity on small length scales.

The fractal repeller is the support of an invariant probability measure
which can be constructed by considering statistical averages
over all the trajectories which have not yet escaped during a time
interval $[-T/2,+T/2]$ and, thereafter, taking the limit
$T\to\infty$.
In this limit the trajectories which do not escape during $[0,+T/2]$ have
initial conditions on the stable manifold
of the repeller, $W_{\rm s}({\mathcal F})$.
On the other hand, the initial conditions of the trajectories which,
in the reverse time direction
do not escape during $[0,-T/2]$ approach the unstable manifold,
$W_{\rm u}({\mathcal F})$. In the limit $T\to\infty$,
imposing no escape over the whole interval $[-T/2,+T/2]$ selects
trajectories which approach closer and closer the repeller,
given by the intersection
$W_{\rm s}({\mathcal F}) \ \cap \ W_{\rm u}({\mathcal F}) = {\mathcal F}$.
Statistical averages over these selected trajectories define an invariant
measure having the fractal repeller for support.

In systems with time-reversal symmetric collision dynamics, such as the
present one, statistical averages calculated over the aforementioned
invariant measure are equivalent to statistical averages over a
conditionally invariant measure defined by selecting the
trajectories which do not escape during $[0,+T/2]$ only and taking the
limit $T\to\infty$. This conditionally invariant measure
has been constructed in Refs. \cite{GB95,G98,CG01} and we report
the reader to these papers for further detail.  Let us here add
that, in the limit of low densities of sinks,
the system tends to the closed finite
Lorentz gas, whereupon the invariant and
the conditionally invariant measures reduce to the microcanonical
equilibrium measure. In this limit, the escape process stops and the fractal
repeller fills the whole phase space.

\subsubsection{Average Lyapunov exponent of the repeller}

The positive Lyapunov of the repeller can be calculated
by averaging over the $N_T$ particles still in the system after time $T$
according to
\begin{equation}
\lambda= \lim_{T \to \infty} \ \lim_{N_0 \to \infty} \ \frac{1}{T} \
\frac{1}{N_T} \ \sum^{N_T} _{j=1} \ \ln \Lambda_T(\Gamma_0^{(j)}),
\label{deflyap}
\end{equation}
where $\Lambda_T$ at time $T$ is the stretching factor of an initial
perturbation
on the trajectory $\Gamma_0^{(j)}$ of surviving particles \cite{GB95,G98,CG01}.
The average (\ref{deflyap}) is equivalent to an average based on
the aforementioned conditional invariant measure as shown in Refs.
\cite{GB95,G98}.

\begin{figure}[htbp]
\begin{center}
\epsfxsize=6cm
\subfigure[]{\epsfbox{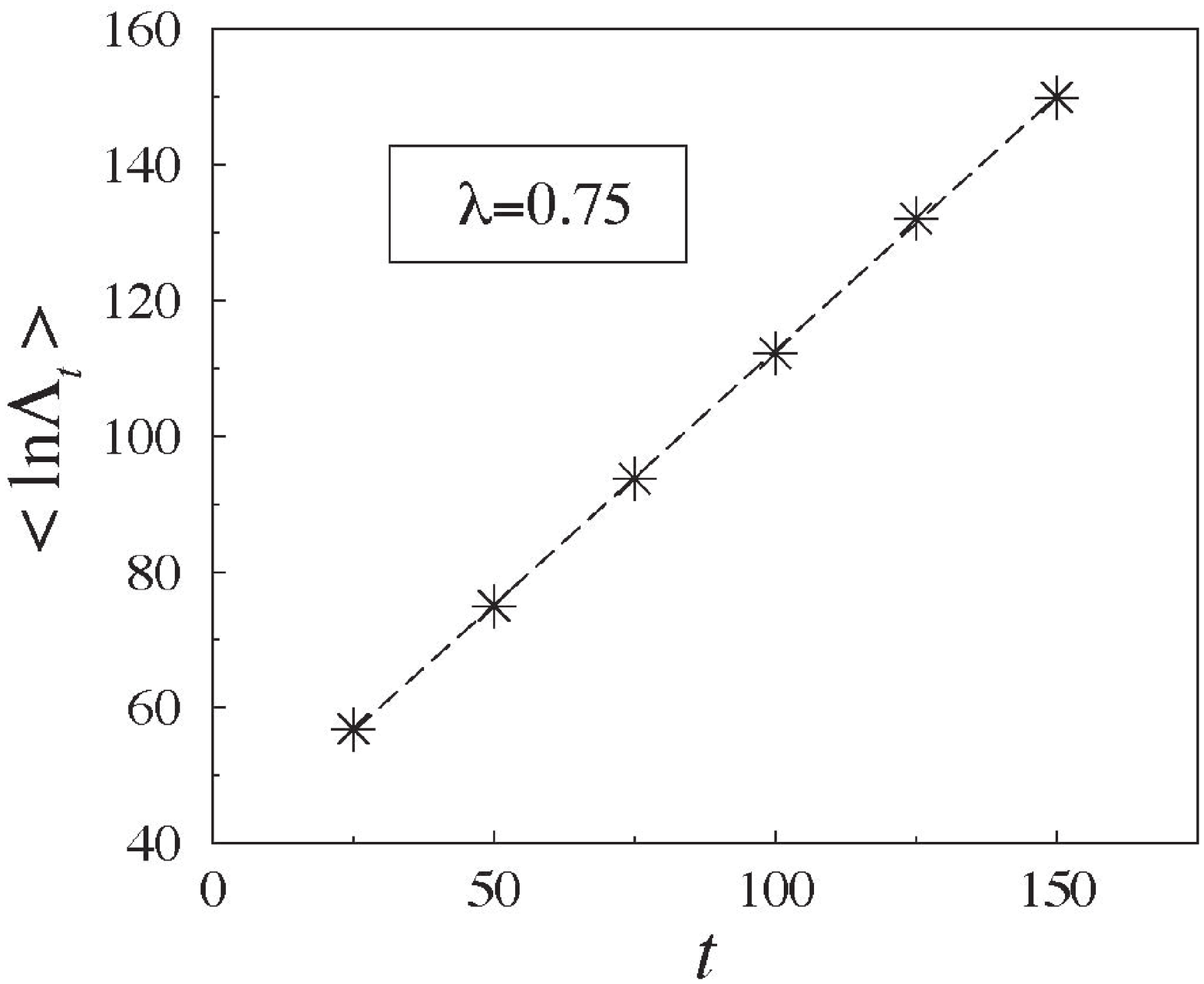}}
\hfill
\epsfxsize=6cm
\subfigure[]{\epsfbox{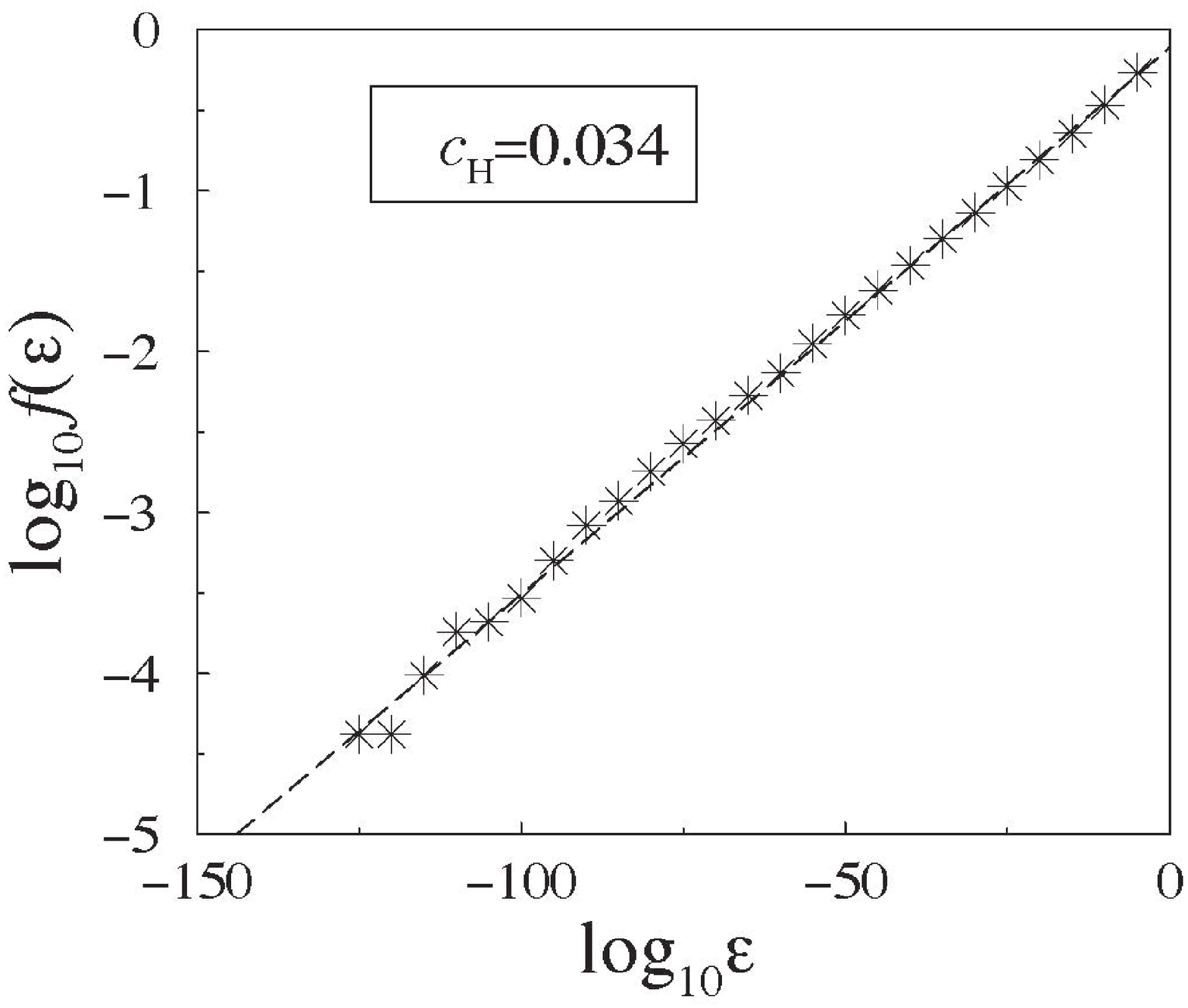}}
\caption{Two-dimensional random Lorentz gas of finite spatial extension
defined by the unit cell represented in Fig.\ \ref{unit12}a:
(a) Average of the logarithm of the stretching factors, as a function of
time
$t$. The slope gives the average Lyapunov exponent $\lambda=0.75$.
(b) Maryland algorithm: the logarithm of the fraction of uncertain pairs
as a function of the logarithm of the difference $\varepsilon$
between the two initial conditions of the same pair. The slope gives the
Hausdorff codimension $c_{\rm H}=0.034$. The product $\lambda c_H=0.025$
is
in good agreement with the value of $\gamma$,
obtained from Fig.\ \ref{esc12}a.
\label{rel1}}
\end{center}
\end{figure}

\begin{figure}[htbp]
\begin{center}
\epsfxsize=6cm
\subfigure[]{\epsfbox{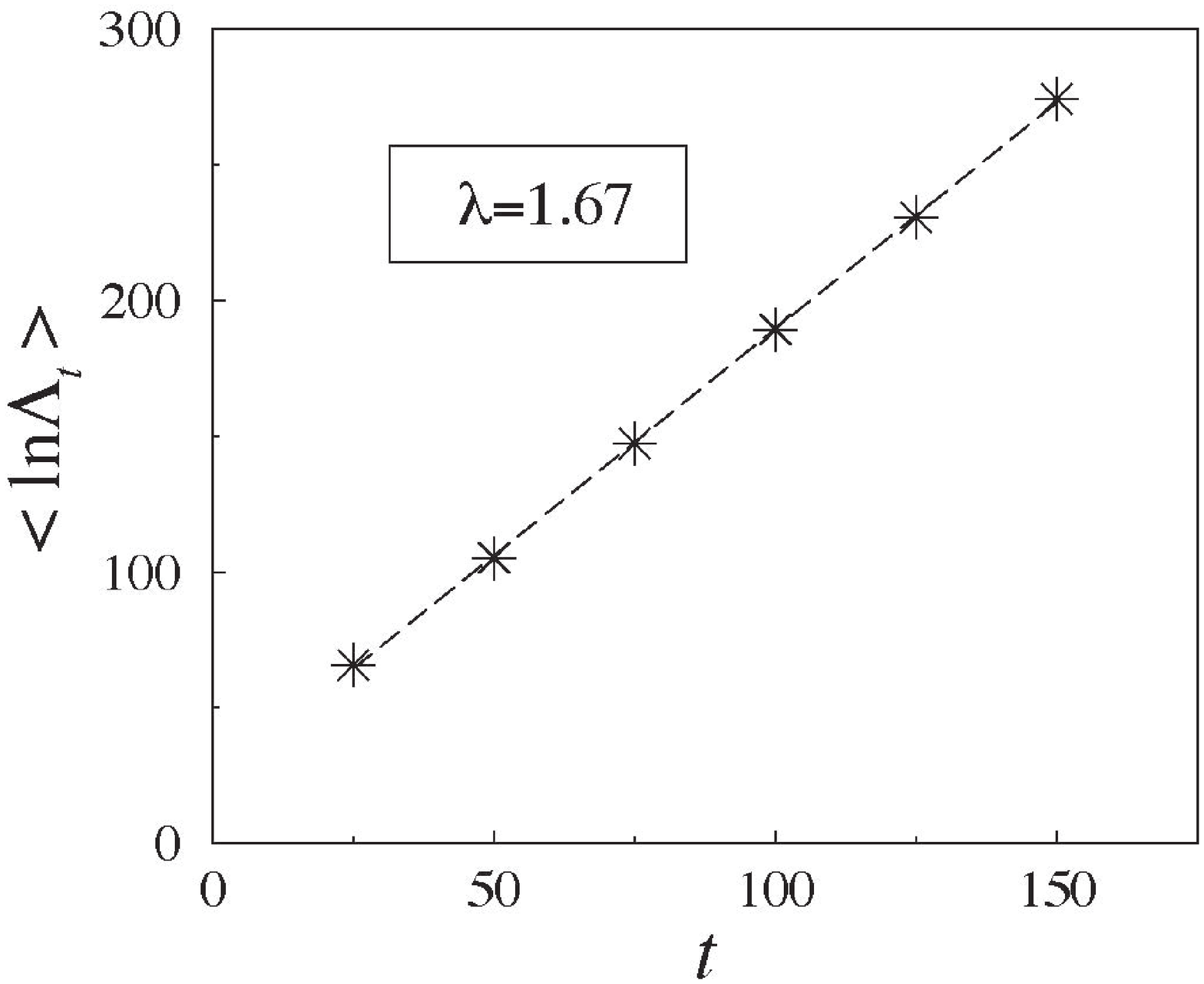}}
\hfill
\epsfxsize=6cm
\subfigure[]{\epsfbox{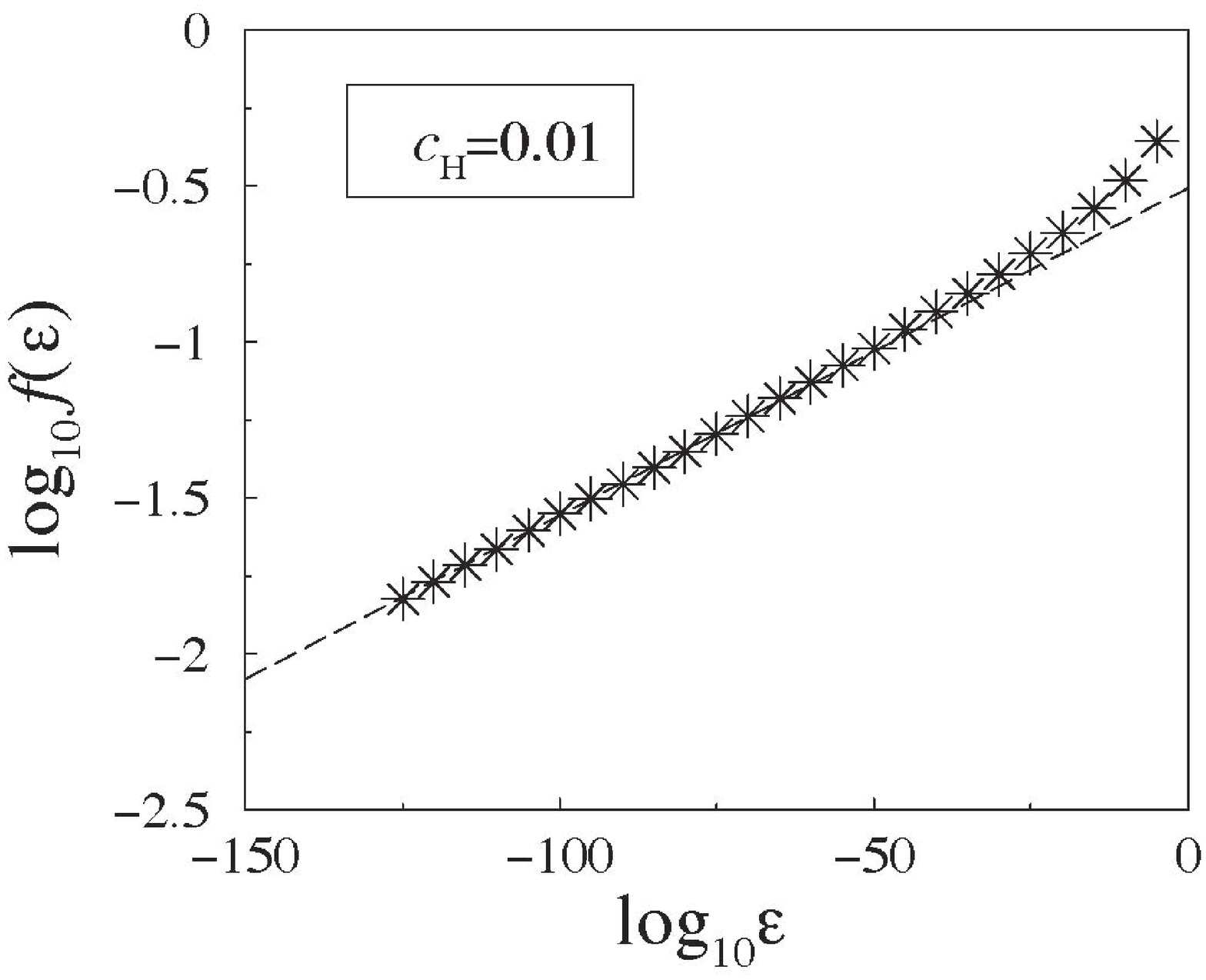}}
\caption{Two-dimensional random Lorentz gas of finite spatial extension
defined by the unit cell represented in Fig.\ \ref{unit12}c:
(a) Average of the logarithm of the stretching factors, as a function of
time
$t$. The slope gives the average Lyapunov exponent $\lambda=1.67$.
(b) Maryland algorithm: the logarithm of the fraction of uncertain pairs
as a function of the logarithm of the difference $\varepsilon$
between the two initial conditions of the same pair. The slope gives the
Hausdorff codimension $c_{\rm H}=0.01$. The product $\lambda c_H=0.017$ is
in good agreement with the value of $\gamma$,
obtained from Fig.\ \ref{esc12}b.
\label{rel2}}
\end{center}
\end{figure}

Numerically, the positive Lyapunov exponent is obtained
as the average slope of the logarithm of the stretching factor as a
function of time according to Eq.\ (\ref{deflyap}).
This is illustrated in Figs.\ \ref{rel1}a and \ref{rel2}a, for the disk and
trap
configurations depicted respectively in Figs.\ \ref{unit12}a and
\ref{unit12}c. The average Lyapunov exponent respectively
equals $\lambda=0.75$ and $\lambda=1.67$.

\subsubsection{Fractal dimensions of the repeller}

Important
characteristics of the fractal properties of
a repeller are
its generalized fractal dimensions.
Repellers most often are multifractals, with
nontrivial generalized fractal dimensions $D_q$ \cite{Hasley}.
Here the embedding dimension of
the repeller is
three, since we are working in a 3-dimensional
phase space. Therefore, the fractal dimensions of the repeller
must satisfy the inequality $0 \le D_q < 3$.
Instead of calculating
directly the dimensions $D_q$ of the fractal repeller of the full flow
we will consider a set of so-called {\it partial} fractal dimensions
$d_q$ defined for the intersection of the stable manifold of the repeller
${\mathcal F}$ with a line $\mathcal L$ across this stable manifold.
Like ${\mathcal F}$ this mostly is a multifractal set. Obviously the
values of $d_q$ must be between 0 and 1.
Because the system is time-reversal symmetric the partial
dimension in the stable direction is equal to the one in
the unstable direction. The direction of the flow
contributes a partial dimension equal to unity.
Hence for given $q$, the dimension $D_q$ is related
to the partial dimension $d_q$ by $D_q=2d_q+1$.

In the present paper, we focus on the fractal dimensions for $q=0$ and $q=1$.
For $q=0$, we have the {\it Hausdorff dimension} commonly defined by
\begin{equation}
d_{\rm H} = d_0=\lim_{\varepsilon \to 0}\ -\frac{\ln N(\varepsilon)}{\ln
\varepsilon} \ ,
\end{equation}
where $N(\varepsilon)$ is the number of intervals $I_j$ of equal size
$\varepsilon$,
required to cover the fractal ${\it f}$ defined on a line.
For $q=1$, we find the so-called {\it information dimension}
\begin{equation}
d_{\rm I} =d_1 = \lim_{\varepsilon \to 0}\ \frac{1}{\ln \varepsilon}
\sum_{j=1}^{N_l} p_j(\varepsilon) \ln p_j(\varepsilon) \ ,
\end{equation}
where $p_j(\varepsilon)=\mu(I_j\cap \it f)$ is the probability weight
of the interval $I_j$ of size $\varepsilon$.
Here, $\mu$ denotes the probability measure describing the distribution of
points in the fractal set $f$.  The Hausdorff and information codimensions
are defined respectively as $c_H=1-d_H$ and $c_I=1-d_I$.

For fractal repellers of an open hyperbolic system,
a relation exists between the partial
information codimension $c_I$, the
(largest) positive Lyapunov exponent
$\lambda$, and the escape rate $\gamma$ \cite{GB95,G98,CG01}
\begin{equation}
\gamma=\lambda \; c_I.
\label{esci}
\end{equation}

To a good approximation the information codimension
can be replaced by the Hausdorff codimension \cite{GB95,G98}.
This last dimension may be calculated with the
aid of a numerical algorithm developed by the group of
Maryland \cite{GB95,G98,McDon}. The basic idea
of this algorithm is to consider an ensemble of pairs of initial conditions,
separated by $\varepsilon$, along the line $\mathcal L$ defined above.
A pair is said to be uncertain if there is a discontinuity of the
escape-time function between both initial conditions. On the other hand,
when the pair is certain, the initial conditions belong to an interval
of continuity of this function.
An estimation of the fraction of uncertain pairs may be
obtained in the following way: typically for an uncertain pair the time after
which the pair becomes uncertain coincides with the time that the
distance between the points becomes of order unity. This time may
thus be estimated as $t_u=(1/\lambda)\ln(1/\varepsilon)$.
Since trajectories escape at the rate $\gamma$,
the fraction of uncertain pairs
would be proportional to $\exp(-\gamma t_u)\sim
\varepsilon^{\gamma/\lambda}=\varepsilon^{c_I}$.
However, such an estimation needs to be refined to take into account the
multifractal character of the repeller and the possibility that $c_I \neq c_H$
due to local fluctuations of the Lyapunov exponents.
A refined estimation of the fraction of uncertain pairs given in
the Appendix \ref{ap} for the simple case of a piecewise-linear
one-dimensional map shows that
\begin{equation}
f(\varepsilon) \sim \varepsilon^{c_{\rm H}} , \label{Mary}
\end{equation}
as proved for the general case in Ref. \cite{McDon}.
We notice that, to leading order in the sink density $c$, the Hausdorff
codimension equals the information codimension \cite{CG01}.

In the case of the Lorentz gas,
discontinuities in the escape-time function are not very practical to
calculate numerically. Instead one may calculate the fraction of pairs that
do not
collide with exactly the same sequence of disks until escape from the
system. For small $\varepsilon$ this will behave in precisely the same way
as the
fraction of uncertain pairs, because it is determined by the same
criterion of reaching a distance of order unity between
both trajectories of a pair before escaping.
The details of the algorithm used here are
described in Refs. \cite{GB95,G98}.

In Figs.\ \ref{rel1}b and \ref{rel2}b, the logarithm of the fraction of
uncertain pairs
(in fact pairs with different collision sequences before escape) is
plotted as a function of the logarithm of the small
$\varepsilon$ separating the two initial conditions of a pair, for the
systems defined respectively by the unit cells depicted in Figs\
\ref{unit12}a and \ref{unit12}c. After a
transient regime, the linear behavior confirms the power-law prediction of
Eq.\ (\ref{Mary}) . The slope gives a value of the Hausdorff
codimension equal to $c_H=0.034$ and $c_H=0.01$ respectively.

For both configurations, the value of the product $\lambda c_H$ is in good
agreement with the value of the corresponding escape rate:
$\lambda c_H=0.025=\gamma$ in the
former case and $\lambda c_H=0.017=\gamma$
in the latter. Since $c_H=c_I$ to leading order in the sink density $c$,
this confirms the validity of the relation (\ref{esci})
for random Lorentz gases of finite spatial extension.
This expression connects directly two quantities characterizing the
microscopic chaotic dynamics of the Lorentz gas, the average Lyapunov
exponent $\lambda$ and the Hausdorff codimension $c_H$ of the fractal
repeller, and a quantity describing the
macroscopic transport behavior of
this open system, the escape rate $\gamma$.

\section
{The two-dimensional random Lorentz gas of infinite spatial extension}
\label{raninf}

\subsection{Description of the model and simulation method}

We now consider a two-dimensional random Lorentz gas of infinite spatial
extension. This supposes that the sinks are randomly distributed among an
infinite number of disks. As a consequence, the density of
larger and larger trap-free regions decays exponentially with their volume
but does not vanish.
A suitable method is needed to simulate this situation.
It cannot be done with the simple method presented in the previous section:
if, in a possibly large, but in practice always finite system,
the nature of the disks is fixed a priori, the size of the largest
trap-free region is determined.
Trap-free regions of a size that is very rare for the given system
size in practice just will not be observed.
However, a certain probability $c$ to be a sink can be
assigned to each disk \cite{An84,Bunde97}.
This leads to an average over the different possible sink configurations,
as a result of which the probabilities for finding large trap-free regions
will attain just the proper values.

Let us define the corresponding configuration-averaged
survival probability $P(c,t)$, where $c$ may be
interpreted alternatively as the average concentration of sinks.
The time evolution of $P(c,t)$ is given by
\begin{equation}
P(c,t)=\sum_n (1-c)^{n}\ p(n,t)=\langle(1-c)^{n_t} \rangle
\label{survprob}
\end{equation}
where $n=n_t$ is the number of {\it different} disks on which the point
particle has collided during the time $t$ and $p(n,t)$ is the probability
that this number is equal to $n$. $\langle \ldots \rangle$ is an ensemble
average. $(1-c)$ is the probability that a disk is not a sink,
so that $(1-c)^{n}$ is the probability that none of the disks met at time
$t$ is a sink, in other words the probability that the point particle
survives.

To compute the probability $p(n,t)$ to meet $n$ different disks
in a time $t$, a purely deterministic simulation is used.
The two-dimensional random Lorentz gas in which the point particle
diffuses is
defined as in the previous section, by a square unit cell of size
$L \times L$. Periodic boundary conditions are applied.
However, an index is associated with each
square cell, allowing us to know which one is currently visited. This is
necessary to distinguish one disk from its copies in the other cells.
Indeed, if the point particle meets a disk and, later, one of its copies, this
should be considered as
collisions with different disks. In other words, although the scatterers
are located on a periodic lattice, be it one with a very large unit cell of
random filling, the distribution of traps has no periodicity
whatsoever.

\subsection{Theoretical predictions and simulation results}

The asymptotic time dependence of the survival probability $P(c,t)$ is
expected to be of the form (\ref{stretch.exp}) \cite{DV,GP,An84,BARK01}.
For checking this dependence in our system, the easiest method
using the idea of Eq.\ (\ref{survprob}) is a direct
simulation, in which $p(n,t)$ is estimated as an average
over a large number of uniformly distributed initial conditions.
The system considered here has as unit cell the one depicted in Fig.\
\ref{unit12}c, of size $25 \times 25$ and with 127 disks. The
number of initial conditions is of the order of $10^8$.

\begin{figure}[htbp]
\centerline{\epsfxsize=8 truecm \epsfbox{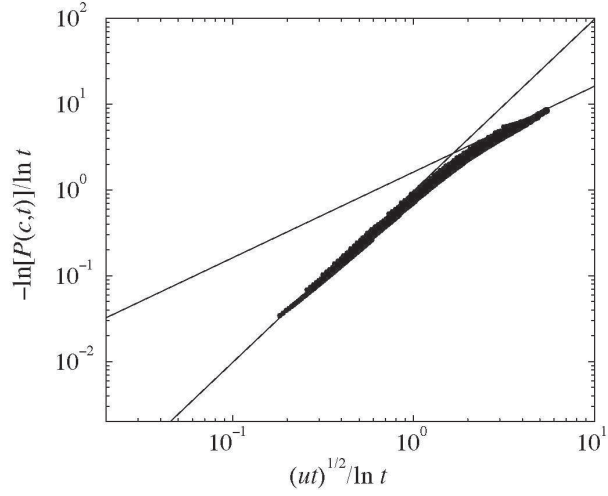}}
\caption{Infinite two-dimensional random Lorentz gas:
Crossover from the Rosenstock regime to the stretched exponential regime.
$-\ln \left[P(c,t) \right]/\ln t$ is plotted as a function of
$\sqrt{ut}/\ln t$ in a double logarithmic plot.
The solid lines are fits to the data with slopes 2 and 1. They cross at
the point $(1.7,2.7)$.
\label{2D}}
\end{figure}

We present our data following a procedure proposed by Barkema
et al. in Ref.\ \cite{BARK01}. At short times and small concentrations, Eq.\
(\ref{survprob}) can be approximated by \cite{ROS}
\begin{equation}
P(c,t)=\langle(1-c)^{n_t} \rangle \simeq (1-c)^{\langle n_t \rangle}.
\end{equation}
This is the so-called Rosenstock approximation. In two dimensions,
$\langle n \rangle$ scales as $4\pi\rho Dt/\ln t$, with $\rho$
the average density of scatterers.
We therefore expect a time dependence of the survival probability of the form:
\begin{eqnarray}
-\ln \left[P(c,t) \right] &\sim& 4\pi\rho Dut/\ln t
\quad
\mbox{at short times and small concentrations}, \nonumber \\
-\ln \left[P(c,t) \right] &\sim& (ut)^{1/2} \qquad \mbox{asymptotically},
\end{eqnarray}
with again $u=-\ln(1-c)$.
An efficient way to visualize the crossover from one regime to the other
is to plot $-\ln\left[P(c,t)\right]/\ln t$
as a function of $\sqrt{ut}/\ln t$ \cite{BARK01,ln8t}.
In the Rosenstock regime, we expect a quadratic dependence, while the
stretched exponential regime should produce a linear behavior. The
results are represented in Fig.\ \ref{2D} in a double-logarithmic plot. The
data
collapse for values of $c$ taken in the range
$[0.01, \ldots,0.09,0.1,\ldots 0.9]$.
The crossover from a slope 2 to a slope 1 is clear.

\begin{figure}[htbp]
\begin{center}
\epsfxsize=6cm
\subfigure[]{\epsfbox{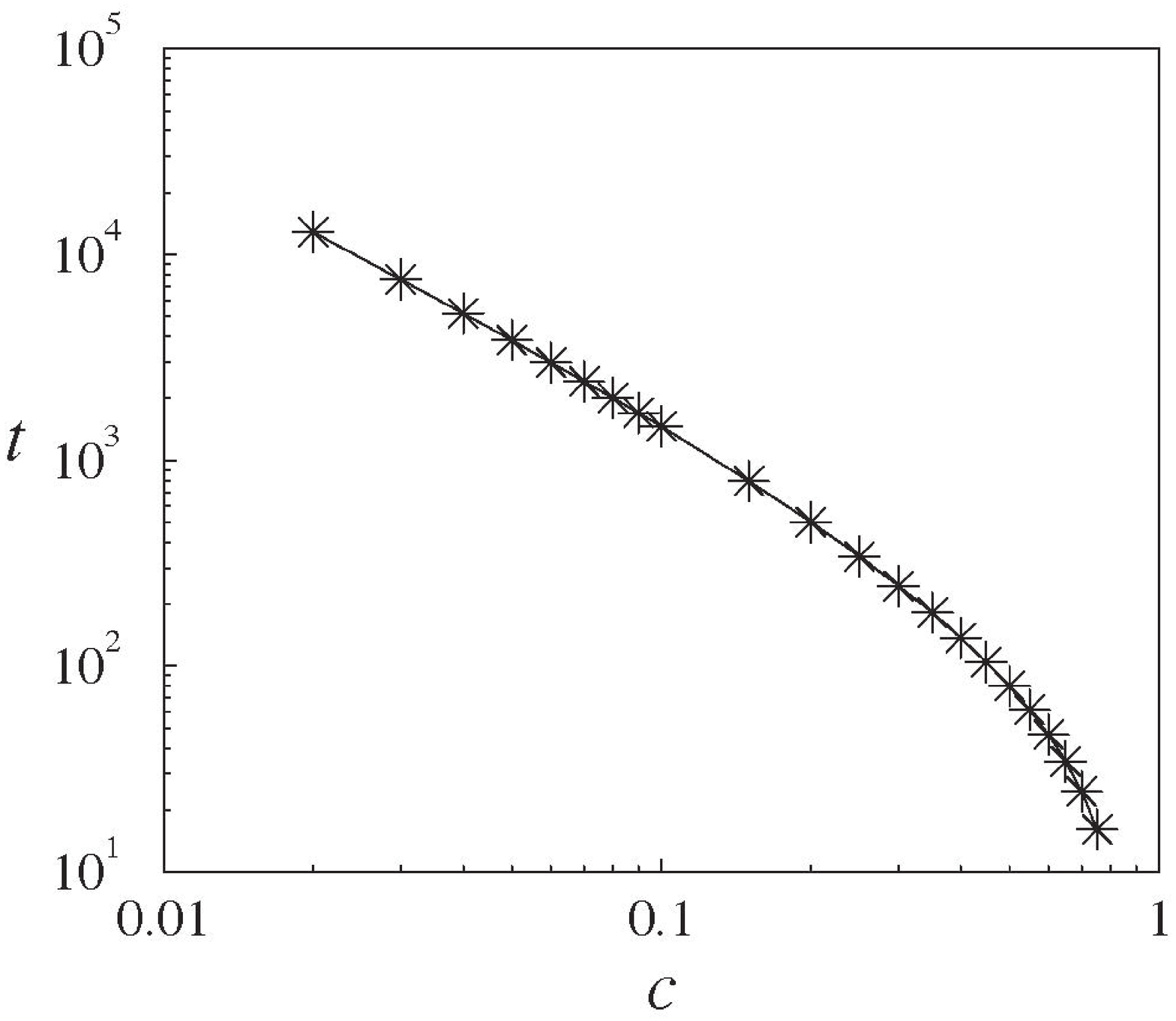}}
\hfill
\epsfxsize=6cm
\subfigure[]{\epsfbox{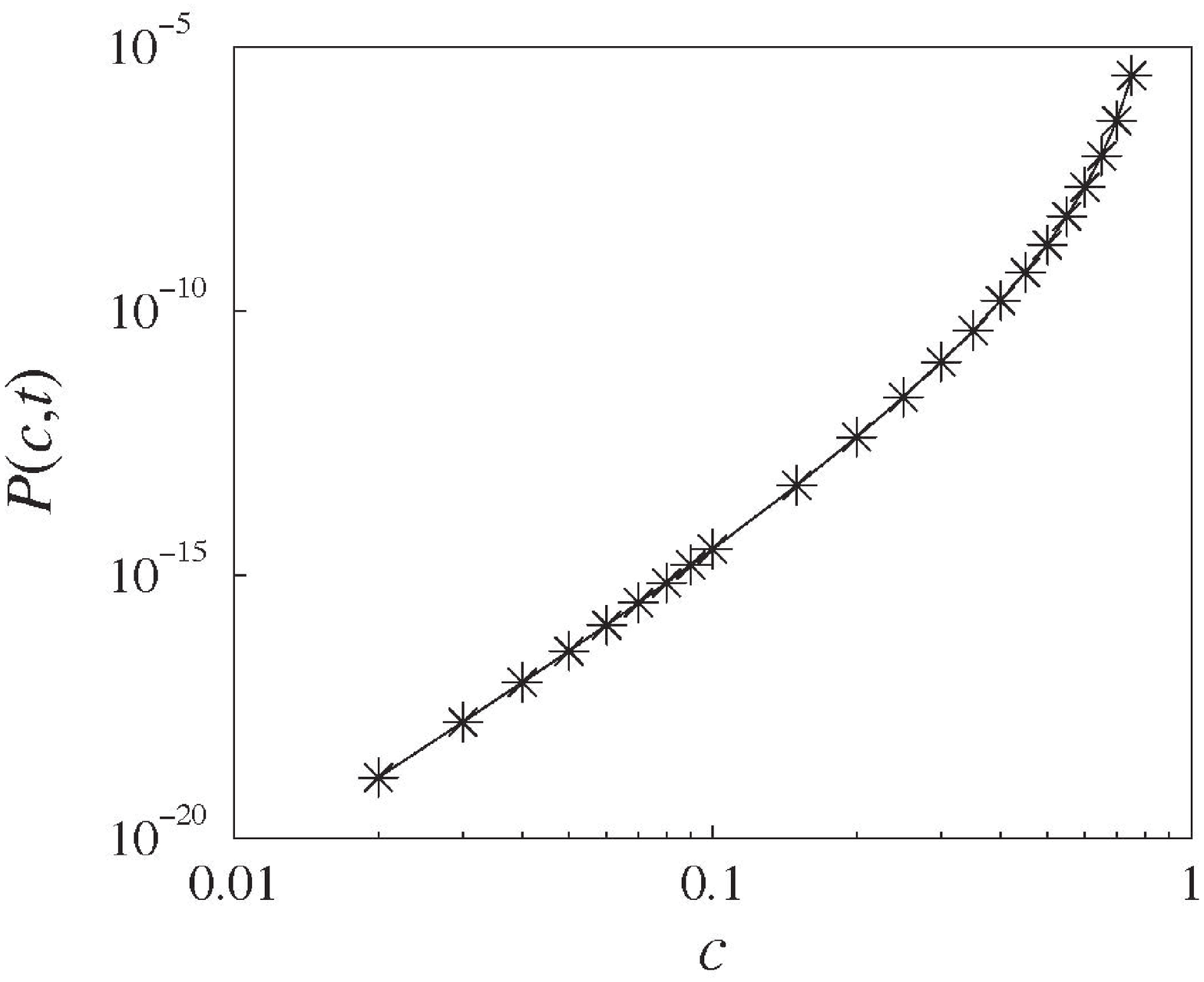}}
\caption{(a) Crossover time from the Rosenstock regime to the stretched
exponential regime as a function of the trap density $c$. (b) Dependence
of the crossover survival probability on the trap density $c$.
\label{cross}}
\end{center}
\end{figure}

The intersection point of the two linear fits gives an estimation
of the crossover time from the Rosenstock regime to another
regime which ends with the Donsker-Varadhan
asymptotics. Using this as reference point, we
obtain the dependence on the trap density $c$ of the crossover time and
of the crossover survival probability \cite{BARK01},
as presented in Figs.\ \ref{cross}a and b respectively.
These results show that the crossover occurs at shorter and shorter times
and at larger and larger probabilities as the trap density increases. We
observe that the crossover probability changes from ${\mathcal O}(10^{-20})$ to
${\mathcal O}(10^{-5})$ as the trap density only varies from
${\mathcal O}(10^{-2})$ to ${\mathcal O}(1)$.

A comment is here in order.  
The extensive numerical simulations of Ref. \cite{MG02}
show that the Donsker-Varadhan behavior is approached only very gradually.
It is known to provide an exact asymptote for the logarithm of $P(c,t)$ 
including the prefactor of $(ut)^{1/2}$. If one inserts the relevant values 
of the parameters one find this asymptote in Fig.\ \ref{2D}
about 20\% above the asymptote obtained from the data fit.
Given the limited time range over which our numerical data were obtained
we may conclude that their behavior is fully compatible
with the one observed for stochastic systems in Ref. \cite{MG02}.

\section{Transition from finite to infinite random systems}
\label{rantrans}

We have considered finite random Lorentz gases with exponential behavior
and infinite ones with stretched exponential behavior.
How does the transition between these two regimes evolve with increasing
system size $L$?
And what happens to the fractal repeller?

To answer these questions, let us first briefly sketch the arguments of
Grassberger and Procaccia for
the infinite case \cite{GP}. They consider diffusing particles distributed
according to a density $\rho_c({\bf r},t)$. The average density
$n_c({\bf r},t)$ is defined as
\begin{equation}
n_c({\bf r},t)=\langle\rho_c({\bf r},t)\rangle_s
\end{equation}
where $\langle...\rangle_s$ is an average over all the realizations
of trap distributions
or equivalently over all choices of the origin from which all particles
start
diffusing initially.
A perfectly absorbing sphere of large radius $R$ centered at the origin is
added to the system. The value of the new density $n'_c({\bf r},t)$
obtained
in the presence of this sphere
is necessarily smaller than $n_c({\bf r},t)$
\begin{equation}
n_c({\bf r},t) \ge n'_c({\bf r},t).
\end{equation}
$n'_c({\bf r},t)$ depends on the distribution of traps inside the sphere.
If $\{S\}=\{S_1,S_2,\ldots,S_k\}$ are the positions of the traps in the
sphere, we have
\begin{equation}
n_c({\bf r},t) \ge n'_c({\bf r},t)=\sum_{\{S\}}n'_c({\bf
r},t|\{S\})\;P(\{S\}).
\end{equation}
The sum of the right-hand side can be bounded from below by considering
only one term, corresponding to the situation of a sphere free of traps:
\begin{equation}
n_c({\bf r},t) > n'_c({\bf r},t|\{S\}= 0)\; P(0).
\label{nc1}
\end{equation}

A random distribution of sinks corresponds to a Poisson distribution. The
probability of finding $k$ traps in a volume $v$ is given by
\begin{equation}
P_k=\frac{\bar N ^k}{k!}\exp(-{\bar N})
\label{pempty}
\end{equation}
with average value ${\bar N}=n_s v$,
where $n_s$ is the average density of sinks.
The probability to have no trap in the sphere is thus given by
$P(0)=\exp(-C\;n_s\;R^d)$, where $d$ is the dimension of the system
and $C$ the volume of the unit ball in $d$ dimensions.

On the other hand, $n'_c({\bf r},t|\{S\}=0)$ is obtained as solution of
a diffusion equation of coefficient ${\mathcal D}_0$
with absorbing boundary conditions on the border
of a sphere of radius $|{\bf r}|=R$. Such a solution is known
to decay asymptotically as the exponential
$\exp(-\kappa_d \frac{{\mathcal D}_0}{R^2}t)$,
where $\kappa_d$ is a constant depending on the dimensionality
of the system.

Therefore, Eq.\ (\ref{nc1}) becomes
\begin{equation}
n_c({\bf r},t) > n'_c({\bf r},t|\{S\}=0) P(0) \sim
\exp\left(-\kappa_d\;\frac{{\mathcal D}_0}{R^2}\;t-Cn_sR^d\right).
\label{nc3}
\end{equation}
The right-hand side may be maximized as a function of $R$,
with $R$ between $0$ and $+\infty$ in an infinite system.
This gives us an optimal value for $R$
\begin{equation}
R=\left( \frac{2 \kappa_d}{d C} \frac{{\mathcal D}_0}{n_s}
t\right)^{1/(d+2)}
=\alpha\; t^{1/(d+2)}.
\label{Rt}
\end{equation}
Inserting this result in Eq.\ (\ref{nc3}), one obtains
\begin{equation}
n_c({\bf r},t) > \exp\left[-\beta t^{d/(d+2)} \right].
\end{equation}
Let us now consider a finite system of linear size $2L$
with a fixed configuration of traps. As long as
$R=\alpha t^{1/d+2} \ll
\left(\frac{\ln L}{n_s}\right)^{1/d}$,
the optimal value of $R$
is found in a fraction of the volume described by (\ref{pempty}) and the
stretched exponential
behavior is observed. However, for longer times, the
maximal value of $R$ is
proportional to
$l_0\approx (\ln L/n_s)^{1/d}$.
Inserting
$R=l_0$ in Eq.\ (\ref{nc3})
leads to an exponential decay of the form
$\exp\left(-\kappa_d\;\frac{{\mathcal D}_0}{
l_0^2}\;t\right)$.
As $L$ increases, the time $t^*$ at which the transition occurs increases.

A logarithmic dependence of $t^*$ on $L$
in two dimensions has been proven in Ref.\
\cite{Bunde97}.

\begin{figure}[htbp]
\centerline{\epsfxsize=4 truecm \epsfbox{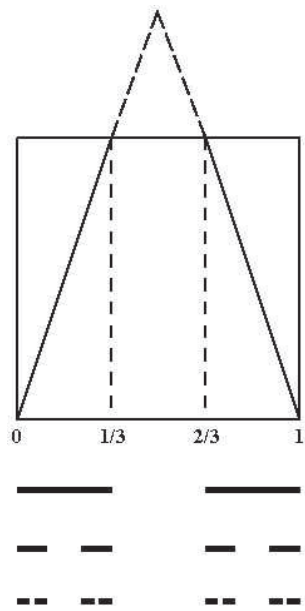}}
\caption{One-dimensional map with escape.
\label{mapesc}}
\end{figure}

We now want to study the properties of the repeller.
As a preparation, let us first
consider the simple one-dimensional map with escape represented in
Fig. \ref{mapesc}.
Starting from the full unit interval at time zero one finds after $k$
discrete time steps an approximate forward repeller consisting of
$N_k=2^k$ pieces of sizes $\varepsilon_k=(1/3)^k$.
The number of pieces is equal to the exponential of the Kolmogorov-Sinai
entropy times time, $N_k=\exp (h_{KS}k)$, and the linear size of these
pieces is given by $\varepsilon_k=\exp(-\lambda k)$ with the
Lyapunov exponent $\lambda=\ln 3$. This last equality can
be intuitively understood as follows: as time goes on, new structures
appear on the repeller on smaller and smaller scales. Therefore,
long times correspond to small spatial scales.
The total length $L_k$ of the repeller at time step $k$ is the product
of $N_k$ and $\varepsilon_k$.
It decreases exponentially with time as
\begin{equation}
L_k = N_k \; \varepsilon_k
=\exp(h_{KS} k)\;\exp(-\lambda k)
= \exp(-\gamma k),
\end{equation}
where $\gamma=\lambda-h_{KS}$ is the escape rate.

The fractal (Hausdorff) dimension of the repeller is thus given by
the well-known result
\begin{equation}
D_H=\lim_{\varepsilon \to 0}\frac{\ln N(\varepsilon)}
{\ln \left(\frac{1}{\varepsilon} \right)}=\frac{h_{KS}}{\lambda}=\frac{\ln2}{\ln3} ,
\end{equation}
in the 
illustrative example of
Fig.\ \ref{mapesc}.
On the other hand, we know that the fraction of
uncertain pairs $f(\varepsilon)$, as defined for the Maryland algorithm,
is proportional to the size $\varepsilon$ to the power $C_H$,
with $C_H$ the codimension of the repeller. From
this we obtain
\begin{equation}
f(\varepsilon) \sim \varepsilon^{C_H} \sim \varepsilon^{1-D_H} \sim
\frac{\varepsilon}{\varepsilon^{D_H}} \sim \varepsilon N(\varepsilon).
\label{fvsN}
\end{equation}

\begin{figure}[htbp]
\begin{center}
\epsfxsize=6cm
\subfigure[]{\epsfbox{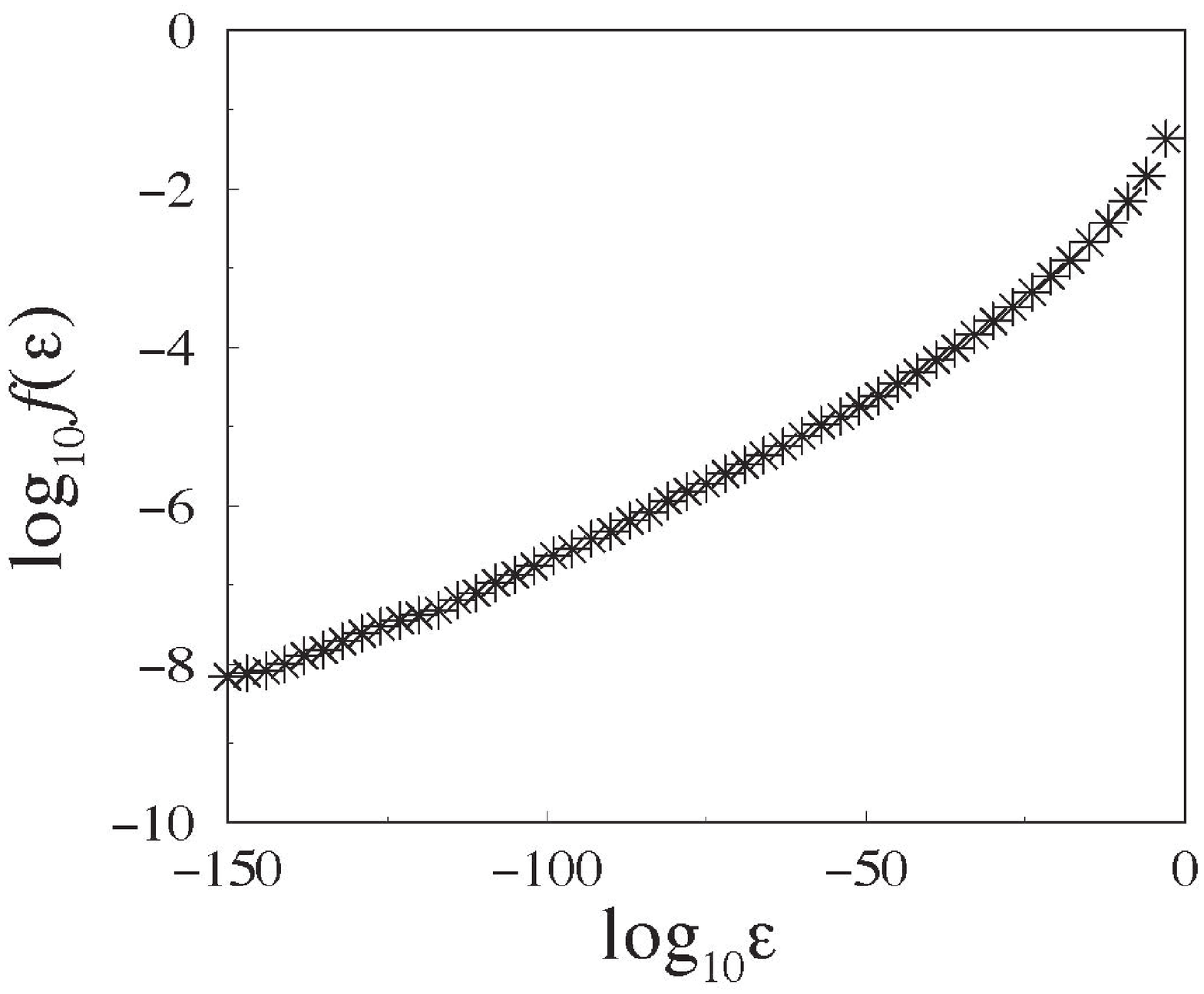}}\\
\epsfxsize=6cm
\subfigure[]{\epsfbox{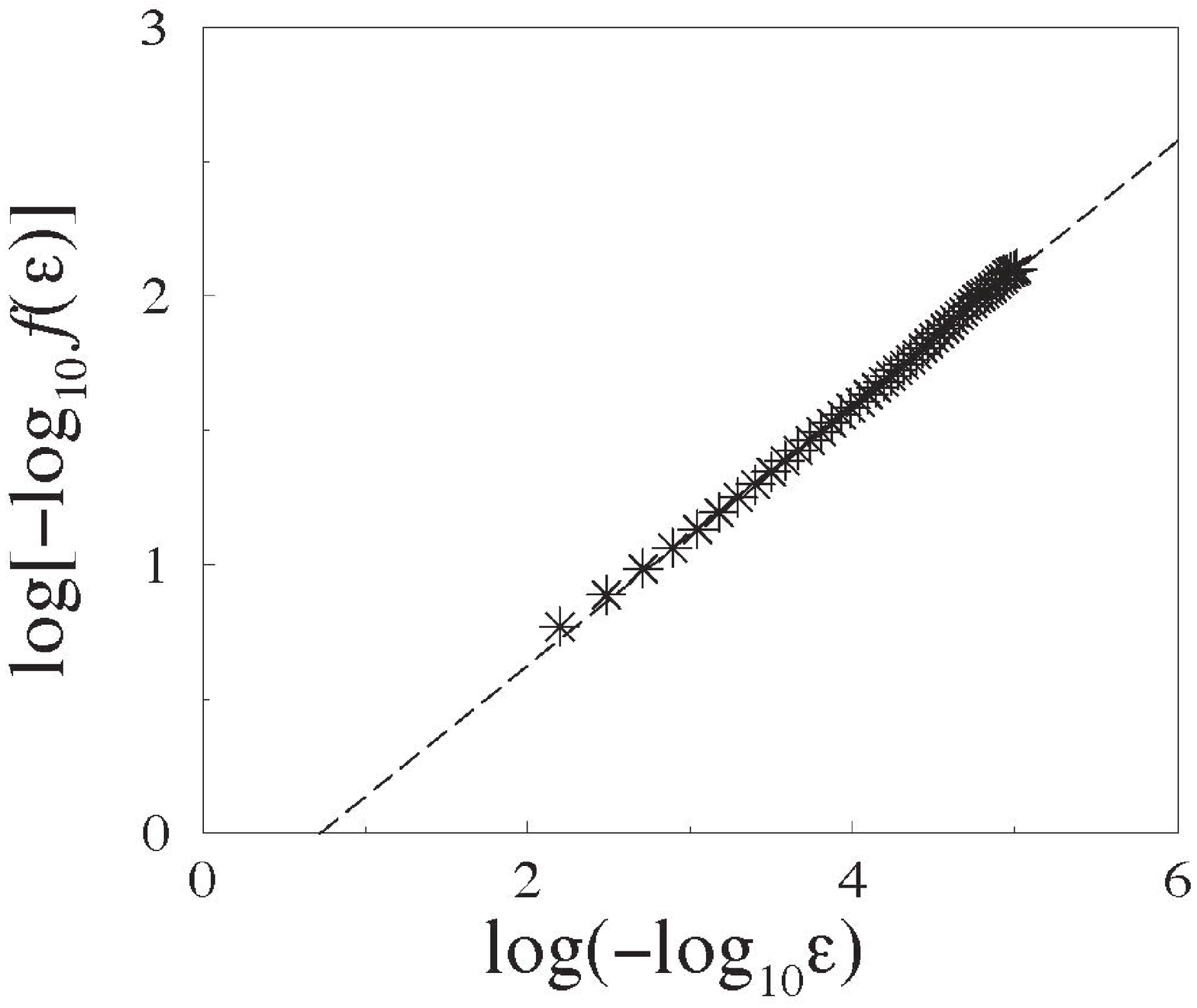}}
\caption{Maryland algorithm in the case of a random Lorentz gas of
infinite spatial extension: the system considered is represented in
Fig.\ \ref{unit12}b, with a sink density $c=0.6$. (a) The logarithm of the
fraction of uncertain
pairs is plotted as a function of the logarithm of the difference
$\varepsilon$ between the two initial conditions of the same pair.
(b) Logarithmic representation of the same quantities,
confirming the dependence predicted by Eq.\ (\ref{fvse}).
\label{hausinf}}
\end{center}
\end{figure}

Let us now imagine a one-dimensional
system in which the escape would obey
a stretched exponential law. In this case, the remaining length of the
repeller at time $k$ would obey
\begin{equation}
L_k=N_k\;\varepsilon_k=\exp(-c\;k^{1/2}).
\end{equation}
>From the relation $\varepsilon=\exp(-k\;\lambda)$, we can express the
time as $k=\frac{1}{\lambda}\;\ln \frac{1}{\varepsilon}$. We get
\begin{equation}
N(\varepsilon) = \frac{L_k}{\varepsilon}=\frac{\exp(-c\;k^{1/2})}
{\varepsilon} = \frac{1}{\varepsilon}\;\exp\left [-c\;\left
(\frac{1}{\lambda}\;
\ln \frac{1}{\varepsilon}\right )^{1/2}\right ].
\end{equation}
>From Eq.\ (\ref{fvsN}), we have
\begin{equation}
f(\varepsilon) \sim \varepsilon N(\varepsilon) \sim \exp \left [ -c\; \left
(\frac{1}{\lambda} \;
\ln \frac{1}{\varepsilon} \right )^{1/2} \right ].
\label{fvse}
\end{equation}
This dependence of the fraction of uncertain pairs on $\varepsilon$
for a system with
stretched exponential behavior can be tested numerically for the
two-dimensio\-nal infinite random Lorentz gas,
with the help of the Maryland algorithm. The
algorithm has to be slighty modified to simulate an infinite system. As
mentioned before, the absorbing or non-absorbing nature of the disks is
not fixed a priori, but each disk has a probability $c$ to be a sink. At each
collision on a disk, the separation of the pair of particles considered
is tested. If  separation occurs, the fraction of uncertain pairs is
increased by $(1-c)^{n}$, where $n$ is the number of {\it different} disks
that the pair has met before separating,
so $(1-c)^{n}$ is the probability that the particles have not escaped
before separating.
Simultaneously, the fraction of escaped pairs is increased by $[1-(1-c)^{n}]$.
In Fig.\ \ref{hausinf}a, the logarithm of the fraction of uncertain pairs,
$\log_{10}f(\varepsilon)$, is plotted as a function of the logarithm of
the distance between the two initial conditions of the pair,
$\log_{10}\varepsilon$. In order to check that this curve can be fitted by
the expression (\ref{fvse}), we
plot in Fig.\ \ref{hausinf}b
$\log[-\log_{10}f(\varepsilon)]$ versus $\log(-\log_{10}\varepsilon)$.
We obtain a linear behavior with
slope $\frac{1}{2}$, in excellent agreement with Eq.\ (\ref{fvse}).

In the case of
a system of finite spatial extension
but with high density of disks,
as represented in Fig.\ \ref{unit12}c,
we see in Fig.\ \ref{esc12}b
that the transient behavior
persists until $t \simeq 50$. According to the relation
$\varepsilon=\exp(t \lambda)$, this corresponds to
$\log_{10} \varepsilon \simeq -34$. This is consistent with the deviation
from the linear curve which is observed in Fig.\ \ref{rel2}b for large values
of $\varepsilon$.

In two-dimensional systems like the Lorentz gases we consider here, the
codimension appearing in the relation
$f(\varepsilon) \sim \varepsilon^{c_H}$ is
a partial codimension, associated with a partial dimension $d_H$. The
total dimension is given by $D_H=1+2d_H$. The total number of small
three-dimensional pieces needed to cover the fractal repeller is given by
\begin{equation}
N(\varepsilon)
\sim\frac{1}{\varepsilon}\left[
\frac{f(\varepsilon)}{\varepsilon}\right]^2
\sim \frac{1}{\varepsilon^3}\exp\left [-2c\;\left (\frac{1}{\lambda}\;
\ln \frac{1}{\varepsilon}\right )^{1/2}\right ].
\label{Nvse}
\end{equation}
This type of dependence of $N(\varepsilon)$ on the size $\varepsilon$
leads
to an integer value of the Hausdorff dimension $D_H$. Indeed,
\begin{equation}
D_H = \lim_{\varepsilon \to 0}\frac{\ln N(\varepsilon)}
{\ln \left(\frac{1}{\varepsilon}\right)} = 3-\lim_{\varepsilon \to
0}2c\left(\frac{1}{\lambda}
\frac{1}{\ln \frac{1}{\varepsilon}}\right)^{\frac{1}{2}}  = 3.
\end{equation}
This dimension does not contain the information concerning the correction
to the power law dependence appearing in Eq.\ (\ref{Nvse}).
We notice that corrections to
a power law also appear for
Brownian motion
in the plane \cite{MAN77}. In this case, the dependence essentially
would take the form
\begin{equation}
N(\varepsilon)\sim\frac{1}{\varepsilon^2} \frac{1}
{\ln \ln \frac{1}{\varepsilon}}
\end{equation}
leading to the integer Hausdorff dimension $D_H=2$.

When the size $L$ of a finite system
with fixed traps increases, the time during which
stretched exponential behavior is observed increases
likewise, and the range of $\varepsilon$ over which
the fraction of uncertain pairs
deviates from a power-law behavior in $\varepsilon$.
Correspondingly, as seen from Fig.\ \ref{hausinf}, the slope in the
linear regime of $\log_{10}f(\varepsilon)$ versus $\log_{10}\varepsilon$
becomes smaller and smaller, i.e., the partial Hausdorff
codimension of the repeller will be smaller. This means that
with increasing $L$
the Hausdorff dimension of the repeller progressively increases
towards the asymptotic value $D_H=3$.
The same holds for the information dimension, as in the dominant regions
the trap density approaches zero and we saw before that both dimensions
approach each other in this limit.

In the case of the two-dimensional random Lorentz gas of infinite
spatial extension, the integer value $D_H=3$ of the dimension of the
fractal repeller can be understood by intuitive arguments. Indeed, this
dimension
corresponds to the case of a closed system, without escape. In an infinite
system, for a given time $t$, there exists with a finite probability a
region free of traps of size $R=\left( \frac{2 \kappa_d}{d C}
\frac{{\mathcal D}_0}{n_s} t\right)^{1/(d+2)}$.
The point particles in this region behave
as particles in a closed system
until they escape through the ``boundary'' at $R$.
As time goes on, larger and larger
trap-free regions have to be considered. Even if their
density decays exponentially with their size,
it does not vanish, so that these regions give the dominant
contribution to the measure of the repeller.

Our results show that the repeller is indeed a fractal object
in spite of its integer dimension $D_H=3$.  This fractality
is characterized by the complicated scaling behavior
of the fraction (\ref{fvse}) of uncertain pairs or of the number (\ref{Nvse})
of pieces needed to cover the fractal.
These theoretical predictions are nicely confirmed by
our numerical results.


\section{Conclusions}\label{concl}

We have studied in this paper several types of two-dimensional random
Lorentz gases with sinks.

The first model considered is a two-dimensional random Lorentz gas with
sinks, of finite spatial extension. In this system, the sinks are randomly
distributed among the disks. Due to the finite size of the system,
the size of the largest trap-free region is bounded.
Therefore, the decay of the number of particles
is asymptotically exponential. For this open system, we
have shown the existence of a fractal repeller formed by the trajectories
that never escape from the system. Using a non-equilibrium measure on
the repeller, we have characterized its chaotic and fractal properties. We
have obtained the escape rate $\gamma$, the
positive Lyapunov exponent $\lambda$, and the
fractal dimensions of the repeller. A relation between $\gamma$,
$\lambda$, and the Hausdorff codimension
$c_H$ of the fractal repeller has been numerically tested
in the limit of low sink densities. This relation
establishes a link between the macroscopic decay and the underlying
microscopic chaos.

The second case considered is the two-dimensional random Lorentz gas
with sinks, of infinite spatial extension. This situation can be simulated
by giving
to each disk a certain probability to be a sink,
while distinguishing between different periodic images of the same disk.
This leads to an average
over the different sink configurations. The decay of the survival
probability asymptotically obeys the stretched exponential law
(\ref{stretch.exp}).
Here this behavior has been studied numerically for the first time
in a deterministic chaotic system.

For a dense 2D random Lorentz gas of finite size $L$, the transient
stretched exponential regime persists during a time fixed in first
approximation by the typical size of the largest trap-free region in the system
[see Eq.\ (\ref{Rt})]. As $L$ increases, the transient regime is observed
for longer and longer times. Simultaneously, the dimension of the
repeller increases, reaching the integer value $D_H=3$ in the
infinite-size limit.  We have here been able to show that
the repeller remains fractal in the limit and to answer the puzzle
mentioned in the introduction by finding the nontrivial scaling
property of this plain fractal.  Its fractal property can be
characterized equivalently by the fraction of uncertain pairs
of trajectories starting from initial conditions separated by $\varepsilon$
or by the number of cells of size $\varepsilon$ needed to cover the fractal.
Both scaling functions show a correction to a simple power law proving
that the repeller remains fractal. The correction depends on the
Lyapunov exponent as well as on the exponent $1/2$ of the stretched exponential
law (\ref{stretch.exp}) in $d=2$.  This nontrivial scaling
is remarkably confirmed by our numerical results.

In conclusion, we have been able to show that, in
a random deterministic chaotic system of infinite spatial extension
with sinks, the repeller composed of all the trajectories that never meet a
sink
is a plain fractal with special scaling properties related
to the stretched exponential decay of the survival probability.


\section*{Acknowledgments}
The authors thank Professor G.\ Nicolis for
support and encouragement in this research. IC and PG are
supported financially by the National Fund for Scientific Research
(F.\ N.\ R.\ S.\ Belgium). H.v.B.\ also acknowledges support
by the Mathematical physics program of FOM and NWO/GBE.


\begin{appendix}
\section{Direct proof that the uncertainty exponent is the Hausdorff
codimension}
\label{ap}

Let us consider the piecewise-linear one-dimensional map:
\be
x_{n+1} = \cases{ \Lambda_0 x_n \qquad 0 < x_n < A \cr
\Lambda_1 (1- x_n) \qquad A < x_n < 1 \cr}
\label{example}
\ee
We assign the symbol $\omega=0$ to the interval $[0,A]$
and the symbol $\omega=1$ to the interval $[A,1]$.

If $\Lambda_0,\Lambda_1>2$, almost all the trajectories escape
after a finite number of iterations.
The subinterval $I_{\omega_0\omega_1...\omega_{n-1}}$ is defined as the
set of
initial conditions for
the trajectories which escape in $n$ iterations and successively
visit the intervals corresponding to the symbolic sequence
$\omega_0\omega_1...\omega_{n-1}$.
These subintervals form a partition of the unit interval $[0,1]$:
\be
\cup_{\omega_0\omega_1...\omega_{n-1}}
I_{\omega_0\omega_1...\omega_{n-1}}=[0,1]
\ee
The length of the subinterval $I_{\omega_0\omega_1...\omega_{n-1}}$ is
\be
\ell_{\omega_0\omega_1...\omega_{n-1}} =
\frac{K}{\Lambda_{\omega_0}\Lambda_{\omega_1}
...\Lambda_{\omega_{n-1}}}
\ee
Since the sum of lengths should be equal to one
\be
\sum_{n=1}^{\infty} \sum_{\omega_0\omega_1...\omega_{n-1}}
\ell_{\omega_0\omega_1...\omega_{n-1}} = 1
\ee
the constant $K$ is given by
\be
K =
\frac{1-\Lambda_0^{-1}-\Lambda_1^{-1}}{\Lambda_0^{-1}+\Lambda_1^{-1}}
\ee
and the previously introduced constant $A$ can now be fixed to the
value:
\be
A = \frac{1 + K}{\Lambda_0}.
\ee
A pair is uncertain if it falls in a subinterval which is smaller than
$\ve$.
Therefore, the fraction of uncertain pairs is equal to
\be
f(\ve) = \sum_{n=1}^{\infty} \sum_{\omega_0\omega_1...\omega_{n-1}}
\ell_{\omega_0\omega_1...\omega_{n-1}}
\Big\vert_{\ell_{\omega_0\omega_1...\omega_{n-1}} < \ve}.
\ee
The sum can be reordered in terms of the numbers $n_0$ (resp. $n_1$)
of
symbol $0$ (resp. $1$) in the
symbolic sequence $\omega_0\omega_1...\omega_{n-1}$. We have that
$n_0+n_1=n$. The fraction of uncertain
pairs becomes
\be
f(\ve) = \sum_{n_0=0}^{\infty} \sum_{n_1=0}^{\infty}
\frac{(n_0+n_1)!}{n_0! \; n_1!}
\frac{K}{\Lambda_0^{n_0}\Lambda_1^{n_1}}
\Big\vert_{n_0\ln\Lambda_0+n_1\ln\Lambda_1 > \ln(K/\ve)}.
\ee
We use the steepest-descent method for this double sum. The domain of
summation is
the set of pairs of non-negative integers $(n_0,n_1)$ above the
straight line
$n_0\ln\Lambda_0+n_1\ln\Lambda_1 = \ln(K/\ve)$ in the plane
$(n_0,n_1)$.
The function to
be summed decreases
exponentially to zero as the exponents $n_0$ and $n_1$ tend to
infinity above the straight line.
Therefore, the sum will be
dominated by the maximum of the function on the straight line.
If we introduce the alternative variables
\be
x_0 \equiv \frac{n_0}{\ln(K/\ve)} \qquad \mbox{and} \qquad x_1 \equiv
\frac{n_1}{\ln(K/\ve)}
\ee
we thus find
\be
f(\ve) \sim \ve \left( \frac{K}{\ve}\right)^D \sim \ve^{1-D}
\ee
with
\begin{equation}
	 D =x_0 \ln
\frac{x_0+x_1}{x_0} + x_1
\ln \frac{x_0+x_1}{x_1}
\label{D}
\end{equation}
and the constraint
\be
x_0\ln\Lambda_0+x_1\ln\Lambda_1 = 1.
\label{C}
\ee
The maximum of the function is given by
the steepest-descent condition
\be
\frac{dD}{dx_0}= \ln \frac{x_0+x_1}{x_0} + \frac{dx_1}{dx_0} \ln
\frac{x_0+x_1}{x_1} = 0.
\label{steep2}
\ee
With the constraint (\ref{C}),
we find that the steepest-descent condition reduces to
\be
\ln\Lambda_1 \ \ln \frac{x_0+x_1}{x_0} = \ln\Lambda_0 \ \ln
\frac{x_0+x_1}{x_1}.
\label{SD}
\ee

We now show that the exponent $D$ is equal to the Hausdorff dimension.
In the present piecewise-linear map, the Hausdorff dimension of the
repeller must
satisfy by definition the following condition
\be
\frac{1}{\Lambda_0^D} + \frac{1}{\Lambda_1^D} = 1.
\ee
Let us show that it is indeed the case. For this purpose,
we first express the quantity $D$
in terms of $x_0$ and $x_1$ by
Eq. (\ref{D}). Then, we use the steepest-descent condition (\ref{SD})
and, thereafter, the constraint
(\ref{C}). The calculations are the following:
\begin{eqnarray}
\frac{1}{\Lambda_0^D} + \frac{1}{\Lambda_1^D}
&=& e^{-x_0 \ln \frac{x_0+x_1}{x_0} \ln\Lambda_0 - x_1
\ln \frac{x_0+x_1}{x_1} \ln\Lambda_0} \nonumber \\
	 & &+ e^{-x_0 \ln
\frac{x_0+x_1}{x_0}
\ln\Lambda_1 - x_1
\ln \frac{x_0+x_1}{x_1} \ln\Lambda_1} \\
&=& e^{-(x_0 \ln\Lambda_0 + x_1 \ln\Lambda_1)\ln \frac{x_0+x_1}{x_0}}
+ e^{-(x_0 \ln\Lambda_0 + x_1
\ln\Lambda_1) \ln \frac{x_0+x_1}{x_1}} \\
&=& e^{-\ln \frac{x_0+x_1}{x_0}} + e^{- \ln \frac{x_0+x_1}{x_1}} = 1
\end{eqnarray}
Q.E.D.

The previous calculation provides a direct proof that the uncertainty
exponent of the
Maryland algorithm is the Hausdorff codimension in a piecewise-linear map
with two branches. Consequently, the uncertainty exponent should be
expected to differ from the information codimension on general
multifractal repellers. Indeed, by using the example (\ref{example}), we
have numerically observed that the uncertainty exponent agrees with the
Hausdorff codimension but differs from the information codimension if
$\Lambda_0\neq\Lambda_1$.

The direct proof presented here can be extended to piecewise-linear
one-dimensional
maps with more than two branches. In the case of $m$ branches, there would
be $m$ variables $n_i$ or $x_i$ and $(m-1)$ steepest-descent conditions such as
(\ref{steep2}) because of the constraint that the maximum point
belongs to the hyperplane $\sum_{i=1}^m n_i\ln\Lambda_i=\ln(K/\varepsilon)$.
Otherwise, a Lagrange multiplier may be introduced,
to deal with this constraint in the search of the maximum. In the
general case, the argument of Ref. \cite{McDon} provides the proof.
\end{appendix}

\newpage

\begin{thebibliography}{10}
\bibitem{ln8t} According to the recent Ref. \cite{MG02}, improved agreement
with the asymptotic formulas can be obtained by including non-leading
corrections.
\end{thebibliography}
\end{document}